\documentclass[aps,prl,twocolumn,superscriptaddress,showpacs]{revtex4}

\usepackage{amsmath,amssymb}
\usepackage{graphicx,dcolumn,bm}

\newcommand{\Tr}{\mathop{\mathrm{Tr}}}
\newcommand{\diff}{\mathrm{diff}}
\newcommand{\opt}{\mathrm{(opt)}}
\newcommand{\abar}{\bar{a}}
\newcommand{\Omegabar}{\bar{\Omega}}
\newcommand{\mubar}{\bar{\mu}}
\newcommand{\nubar}{\bar{\nu}}
\newcommand{\Tcal}{{\cal T}}
\newcommand{\Kcal}{{\cal K}}

\newcommand{\Mcal}{{\cal M}}
\newcommand{\Tzs}{T_{\Delta^*}^z}

\newcommand{\bra}[1]{\langle #1|}
\newcommand{\ket}[1]{|#1 \rangle}
\newcommand{\bracket}[2]{\langle #1|#2 \rangle}

\begin{document}


\title{Systematic Derivation of Order Parameters through 
Reduced Density Matrices}



\author{Shunsuke Furukawa}
\affiliation{Department of Physics, Tokyo Institute of Technology, 
Meguro-ku, Tokyo 152-8551, Japan}
\author{Gr\'egoire Misguich}
\affiliation{Service de Physique Th\'eorique, CEA Saclay, 
91191 Gif-sur-Yvette Cedex, France}
\author{Masaki Oshikawa}
\affiliation{Department of Physics, Tokyo Institute of Technology, 
Meguro-ku, Tokyo 152-8551, Japan}


\date{\today}

\begin{abstract}
A systematic method for determining order parameters 
for quantum many-body systems on lattices is developed
by utilizing reduced density matrices.  
This method allows one to extract the order parameter 
directly from the wave functions of the degenerate ground states 
without aid of empirical knowledge, 
and thus opens a way to explore unknown exotic orders.  
The applicability of this method is demonstrated 
numerically or rigorously 
in models which are considered to exhibit 
dimer, scalar chiral, and topological orders.  
\end{abstract}


\maketitle


Determining order parameters is one of the most important issues 
in the study of many-body systems.  
A suitably chosen order parameter for a symmetry-breaking phase 
provides an intuitive picture of the long range order 
and is the necessary starting point of the 
Landau-Ginzburg-type effective description of the system \cite{Landau}.  
Combined with Wilson's idea of renormalization group \cite{Wilson}, 
such an effective theory becomes a powerful tool in analyzing 
the nature of the phase transitions to other phases.  

In spite of the importance of this issue, 
a general method to obtain an order parameter in a given model 
is not available.  
The knowledge of previous examples suggests some candidates 
but this empirical method may fail in the case of a new order.  
Especially in a system 
with strong frustration and/or quantum fluctuation, 
the order parameter can be quite non-trivial.
With more examples of exotic orders becoming
a subject of great theoretical and experimental interest,  
a {\em systematic} method for determining an order parameter 
would be strongly desired.

In this letter, we present a solution 
to the quantum version of this problem.  
In the quantum case, 
when a symmetry of the Hamiltonian is broken spontaneously
in the thermodynamic limit, 
there appear degenerate ground states (GS). 
An order parameter can be identified with an operator 
which distinguishes the degenerate GSs.  
The central idea of our method is to search such an operator 
by comparing the reduced density matrices (RDM) of 
the degenerate GSs for various subareas of the system.  
A RDM efficiently encapsulates the expectation values of 
all the operators on the concerned area.  
If the RDMs of the GSs are different on an area, 
an order parameter can be defined on that area.  
In this way, we can determine the smallest area 
on which an order parameter can be defined.  
Moreover, for the resultant area, 
we can construct an ``optimal'' order parameter from the RDMs.  
This method can be applied to the low-energy eigenstates 
obtained by exact diagonalization, for instance, 
and can reveal the order parameter without bias.

We will demonstrate the effectiveness of this approach 
in concrete models.  
We will consider the multiple-spin exchange model on the ladder 
and detect dimer and scalar chiral orders 
which have been found in previous studies \cite{Laeuchli,Hikihara}.  
We will also consider a resonating valence bond (RVB) liquid 
in a solvable quantum dimer model (QDM) \cite{Misguich_kagome} 
and rigorously show that its GSs cannot be 
characterized by {\em any} local order parameter \cite{note_local}.
Namely, the model will unambiguously be shown 
to posses a topological order.

{\em Methodology ---}
Suppose that we have obtained the low-energy spectrum and eigenstates 
of finite-size systems by exact diagonalization, for instance.  
In a phase breaking a {\em discrete} symmetry, 
we find a finite number of nearly-degenerate GSs 
which become asymptotically degenerate 
when increasing the system size.  
Each of these states does not break any symmetry of the Hamiltonian 
but its quantum numbers indicate 
what symmetries are broken in the thermodynamic limit.  

Let us focus on the simplest case:  
the Hamiltonian is invariant 
under the translation by one lattice spacing, $\Tcal$, 
is real in terms of $\{S_j^z\}$-basis, 
and exhibits doubly-degenerate GSs, 
$\ket{\Phi_1}$ and $\ket{\Phi_2}$, 
with momenta $k=0$ and $\pi$, respectively.  
In this case, we expect 
the breaking of the translational symmetry 
(doubling of the unit cell) 
in the thermodynamic limit.  
We set $\ket{\Phi_1}$ and $\ket{\Phi_2}$ real, i.e., 
$\Kcal\ket{\Phi_i}=\ket{\Phi_i}~(i=1,2)$, 
where $\Kcal$ denotes the time-reversal operator 
which converts every component of a vector 
into its complex conjugate in terms of $\{S_j^z\}$-basis 
\cite{note_timereversal}.  

We construct the symmetry-breaking GSs, 
$\ket{\Psi_1}$ and $\ket{\Psi_2}$, 
as linear combinations of 
$\ket{\Phi_1}$ and $\ket{\Phi_2}$.  
We require that they be orthogonal 
($\bracket{\Psi_1}{\Psi_2}=0$) 
and be exchanged under $\Tcal$ 
($\Tcal\ket{\Psi_{1(2)}} \propto \ket{\Psi_{2(1)}}$).  
There are still two possibilities, depending on 
whether the time-reversal symmetry is broken 
($\Kcal\ket{\Psi_{1(2)}} \propto \ket{\Psi_{2(1)}}$) 
or not ($\Kcal\ket{\Psi_{1(2)}} \propto \ket{\Psi_{1(2)}}$).  
For each case, the symmetry-breaking GSs are constructed as 
\begin{eqnarray}
\!\!|\Psi_{1,2}\rangle&=&(|\Phi_1\rangle\pm|\Phi_2\rangle)\big/\sqrt{2} 
\quad\text{(${\cal K}$-unbreaking case)}, \nonumber \\
\!\!|\Psi_{1,2}\rangle&=&(|\Phi_1\rangle\pm i|\Phi_2\rangle)\big/\sqrt{2}
\quad\text{(${\cal K}$-breaking case)}.
\label{eq:GSs}
\end{eqnarray}
Here the both possibilities have to be examined 
since, due to the anti-unitarity of the time-reversal operator, 
we do not know from the quantum numbers 
whether the system breaks the time-reversal symmetry or not.    

Next we search an operator which distinguishes 
the symmetry-breaking GSs by comparing the RDMs 
$\rho_\Omega^i=\Tr_{\bar{\Omega}} |\Psi_i\rangle\langle\Psi_i|~(i=1,2)$, 
where $\Omega$ is an area in the system and $\bar{\Omega}$ its complement.  
To quantify to what extent the two RDMs are different, 
we introduce a measure as 
\begin{equation}\label{diff-def}
 \diff(\rho_\Omega^1,\rho_\Omega^2)\equiv \max_{|{\cal O}_\Omega|\le 1}
 \bigg| \Tr_\Omega ({\cal O}_\Omega \rho_\Omega^1)
       -\Tr_\Omega ({\cal O}_\Omega \rho_\Omega^2)\bigg|,
\end{equation}
where ${\cal O}_\Omega$ is a variational (hermitian) operator on $\Omega$ 
satisfying $|\langle\psi|{\cal O}_\Omega|\psi\rangle|\le 1$ 
for any normalized vector $|\psi\rangle$. 
If $\diff (\rho_\Omega^1,\rho_\Omega^2)>0$, 
there exists an operator on $\Omega$ 
distinguishing $|\Psi_1\rangle$ and $|\Psi_2\rangle$.   
This measure has the following useful properties.  
a) normalization to a definite range 
$0\le\diff(\rho_\Omega^1,\rho_\Omega^2)\le2$, 
for an arbitrary area $\Omega$.  
b) monotonicity: 
if an area $\Lambda$ completely contains an area $\Omega$, we have 
$\diff(\rho_\Omega^1,\rho_\Omega^2)
\le\diff(\rho_\Lambda^1,\rho_\Lambda^2)$.  

Using the eigenvalues $\{\lambda_j\}$ 
and the eigenvectors $\{|j\rangle\}$ of 
$\Delta\rho_\Omega\equiv\rho_\Omega^1-\rho_\Omega^2$, 
Eq. \eqref{diff-def} can be simplified as 
\begin{eqnarray}
 \diff(\rho_\Omega^1,\rho_\Omega^2)
 = \max_{|{\cal O}_\Omega|\le1} 
      \bigg|\sum_j\lambda_j\langle j|{\cal O}_\Omega|j\rangle \bigg|
 = \sum_j |\lambda_j|.  
\end{eqnarray}
Here the maximization is done by the ``optimal'' order parameter: 
\begin{equation}
 {\cal O}_\Omega^\mathrm{(opt)}
 =\sum_j |j\rangle\ \mathrm{sgn}~\lambda_j\ \langle j|, 
\label{eq:optimal}
\end{equation}
where $\mathrm{sgn}~\lambda_j$ is the sign of $\lambda_j$ 
if $\lambda_j\ne0$ and is zero if $\lambda_j=0$.  
Both the measure and the optimal order parameter can be calculated 
by (numerically) diagonalizing $\Delta\rho_\Omega$.  
As we have discussed above, generally we have to
examine both the $\Kcal$-unbreaking and $\Kcal$-breaking
combinations in eq.~(\ref{eq:GSs}).
In the following, we denote the measure~(\ref{diff-def})
for the $\Kcal$-unbreaking and breaking cases as
``diff1'' and ``diff2'', respectively.

The generalization of this method to systems 
with more than two degenerate GSs 
can be formulated as an optimization problem, 
which will be presented elsewhere.



{\em Simple examples ---}  
To illustrate this method, 
let us consider two simple examples, 
N\'eel and dimer orders.
The corresponding symmetry-unbreaking GSs are respectively given by 
\begin{eqnarray}
 |\Psi_{1,2}^{\text{N\'eel}}\rangle&\!\!\!=\!\!\!&\frac1{\sqrt{2}} 
 (|\uparrow\downarrow\cdots\rangle \pm 
  |\downarrow\uparrow\cdots\rangle),\\
 |\Psi_{1,2}^{\text{dimer}}\rangle&\!\!\!=\!\!\!&\frac1{\sqrt{2}} 
 (|s_{12}\rangle\cdots|s_{N-1,N}\rangle \pm 
  |s_{23}\rangle\cdots|s_{N,1}\rangle),
\end{eqnarray}
where $|s_{ij}\rangle$ denotes a singlet bond.  
For the N\'eel order, $\mathrm{diff1}=2$
for a 1-site area $\{1\}$ 
while diff2
is zero for the same area.  
Thus the optimal order parameter should be constructed 
from the RDMs of the $\Kcal$-unbreaking GSs on $\{1\}$, 
resulting in ${\cal O}_{\{1\}}^{\opt}=2S_1^z$.
For the dimer order, on the other hand, 
both ``diff1'' and ``diff2'' are zero for $\{1\}$ 
but we find $\mathrm{diff1}=3/2 > 0$ for a 2-site area $\{1,2\}$.  
The resultant optimal order parameter~(\ref{eq:optimal})
for this area is 
${\cal O}_{\{1,2\}}^{\opt}=2\bm{S}_1\cdot\bm{S}_2+1/2$.
We have obtained the expected order parameters 
for both of the simple examples. 


Actually, in order to establish the presence
(or absence) of an order parameter on a given finite area,
the measure ``diff''~(\ref{diff-def})
has to be defined in the thermodynamic limit.
However, in most of the applications, especially in numerical calculations,
we would be only able to calculate the corresponding quantity
in finite systems.
We expect that, in a gapped system,
the ``diff'' should converge exponentially to the
true value, when the system size is taken to infinity.
While the systematic study of such finite-size effect is outside
the scope of the present Letter, the following application demonstrates
the usefulness of our measure even in
numerical diagonalization of relatively small systems.

\begin{figure}[t]
\centerline{\includegraphics[width=\linewidth]{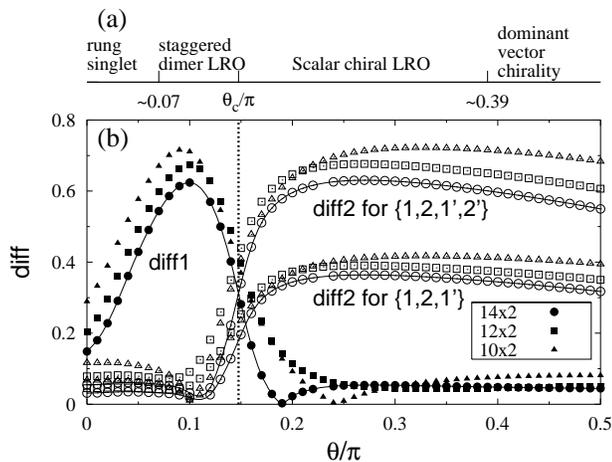}}
\caption{\label{figure_transition}
(a) Phase diagram obtained from earlier studies \cite{Laeuchli,Hikihara}.  
(b) Values of ``diff1'' and ``diff2'' for the fixed areas, 
$\{1,2,1^\prime\}$ and $\{1,2,1^\prime,2^\prime\}$, versus $\theta$.
The values of ``diff1'' (filled symbols) for the two areas 
are exactly the same.  
As for ``diff2'' (open symbols), the upper and the lower points 
refer to $\{1,2,1^\prime,2^\prime\}$ and $\{1,2,1^\prime\}$,
respectively. 
Lines are guides to eyes.  
The vertical dotted straight line represents 
the self-dual point $\theta_c$.  
Our calculation assumes a quasi-degeneracy of the GSs 
and is therefore invalid 
for $\theta\lesssim0.07\pi$ and $0.39\pi\lesssim\theta$.
}
\end{figure}

{\em Application I ---}
We consider the 2-spin and 4-spin exchange model 
with spin $S=1/2$ on the two-leg ladder: 
\begin{equation}
 {\cal H}=\cos\theta\sum_- \bm{S}_i\cdot\bm{S}_j
  +\sin\theta\sum_\square (P_4+P_4^{-1}),
\end{equation}
where the two summations run over 
the (vertical and horizontal) bonds and the squares,
respectively.  
According to recent analyses \cite{Laeuchli,Hikihara}, 
two ordered phases breaking the translational symmetry have been found: 
the staggered dimer phase ($0.07\pi\lesssim\theta <\theta_c$)
and the scalar chiral phase ($\theta_c<\theta\lesssim 0.39\pi$), 
separated by the exact self-dual point 
$\theta_c=\tan^{-1}(1/2)\simeq 0.1476\pi$ \cite{Hikihara}; 
see Fig. \ref{figure_transition}(a).  

\newcounter{tick}
\begin{figure}
\begin{center}
\begin{picture}(140,40)(-10,-10)
  \put(0,10){\line(1,0){120}}
  \put(0,-10){\line(1,0){120}}
  \setcounter{tick}{0}
  \multiput(0,0)(20,0){6}{%
   \put(10,10){\circle*{4}}
   \put(10,-10){\circle*{4}}
   \put(10,-10){\line(0,1){20}}
   \addtocounter{tick}{1}
   \put(7,-20){\arabic{tick}}
   \put(7,15){\arabic{tick}${}^\prime$}
  }
\end{picture}
\end{center}
\caption{\label{figure_ladder}
Numbering of the sites on the two-leg ladder.}
\end{figure}
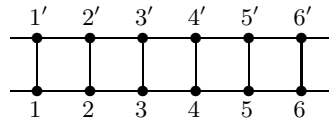

\begin{table}
\begin{tabular}{|l|ll|ll|}
 \hline
 & \multicolumn{2}{c|}{$\theta=0.12\pi$} & 
   \multicolumn{2}{c|}{$\theta=0.19\pi$} \\
 \cline{2-5}
 Area $\Omega$ & diff1 & diff2 & diff1 & diff2\\ \hline
   $\{1\}$                       & $0^*$  & $0^*$  & $0^*$  & $0^*$  \\
   $\{1,2\}$                     & 0.5698 & $0^*$  & 0.0029 & $0^*$  \\
   $\{1,1^\prime \}$             & $0^*$  & $0^*$  & $0^*$  & $0^*$  \\ 
   $\{1,2^\prime\}$              & $0^*$  & $0^*$  & $0^*$  & $0^*$  \\
   $\{1,2,1^\prime\}$            & 0.5698 & 0.0267 & 0.0029 & 0.3340 \\
   $\{1,2,3\}$                   & 0.6579 & 0.0670 & 0.0033 & 0.2365 \\
   $\{1,2,1^\prime,2^\prime\}$   & 0.5698 & 0.0462 & 0.0029 & 0.5785 \\
 \hline
\end{tabular}
\caption{\label{table_diff}
Values of ``diff'' for various areas in the $14\times2$ ladder.  
The points, $\theta=0.12\pi$ and $0.19\pi$, are 
the representative points 
respectively in the staggered dimer and the scalar chiral phases 
found previously \cite{Laeuchli,Hikihara}.  
The sites are numbered as shown in Fig.~\ref{figure_ladder}.  
Some zeros (indicated by $*$) are exact consequences of the symmetries.}
\end{table}

In both regions, the finite-size spectra 
obtained from exact diagonalization exhibit 
two nearly-degenerate singlet GSs with quantum numbers, 
$(k_x,k_y,\sigma)=(0,0,1)$ and $(\pi,\pi,-1)$, 
where $(k_x,k_y)$ denotes the momentum and  
$\sigma$ the reflection with respect to a rung.  
We constructed symmetry-breaking GSs from these states 
and calculated ``diff'' for various areas; see Table \ref{table_diff}.  
At $\theta=0.12\pi$, 
the minimum area required to find an order parameter is $\{1,2\}$ 
and the time-reversal symmetry is unbroken.  
Since $\Delta\rho_{\{1,2\}}$ is symmetric 
under the spin rotations and the time reversal, 
it must be proportional to $\bm{S}_1\cdot\bm{S}_2$ and 
hence the optimal order parameter is 
${\cal O}_{\{1,2\}}^\opt=2\bm{S}_1\cdot\bm{S}_2+1/2$.  
At $\theta=0.19\pi$, the minimum area consists of three sites 
(e.g., $\{1,2,1^\prime\}$) and 
the time-reversal symmetry is broken.  
Since $\Delta\rho_{\{1,2,1^\prime\}}$ 
is symmetric under the spin rotations 
and antisymmetric under the time reversal, we have 
$\Delta\rho_{\{1,2,1^\prime\}}
\propto\bm{S}_1\cdot(\bm{S}_2\times\bm{S}_{1^\prime})$, 
and hence 
${\cal O}_{\{1,2,1^\prime\}}^\opt=\frac{4}{\sqrt{3}}
\bm{S}_1\cdot(\bm{S}_2\times\bm{S}_{1^\prime})$. 
In this way, we have {\em derived} 
the dimer and the scalar chiral operators 
as the appropriate order parameters in a systematic way.  

In Fig. \ref{figure_transition}(b), 
$\theta$-dependence of ``diff1'' and ``diff2'' 
are shown for fixed areas.  
Rapid changes in the values of ``diff'' can be seen 
around the self-dual point $\theta_c$, 
confirming the phase transition between the two ordered phases.
For $\{1,2,1^\prime,2^\prime\}$, 
the values of ``diff1'' and ``diff2'' cross exactly at $\theta_c$.  
For $\{1,2,1^\prime\}$, 
the crossing of ``diff1'' and ``diff2'' 
deviates from $\theta_c$ but approaches it 
when increasing the system size.
In general, such a crossing 
indicates a phase transition between ordered phases 
which cannot be distinguished by the GS quantum numbers.  

{\em Application II ---}
We next consider a solvable QDM 
on the kagome lattice introduced recently \cite{Misguich_kagome} 
(for a review, see also section 5 of Ref. \cite{Misguich_review}), 
which is one of the microscopic models realizing 
a short-ranged (so-called $\mathbb{Z}_2$) RVB liquid.  
This model also provides an example of a solvable Hamiltonian 
\cite{Misguich_qubit}
for a topological quantum-bit based on a QDM \cite{Ioffe_Nature}.  
Before applying our method, 
we briefly review the definition of this model 
and some basic concepts.  

\begin{figure}[t]
\centerline{\includegraphics[width=0.5\linewidth]{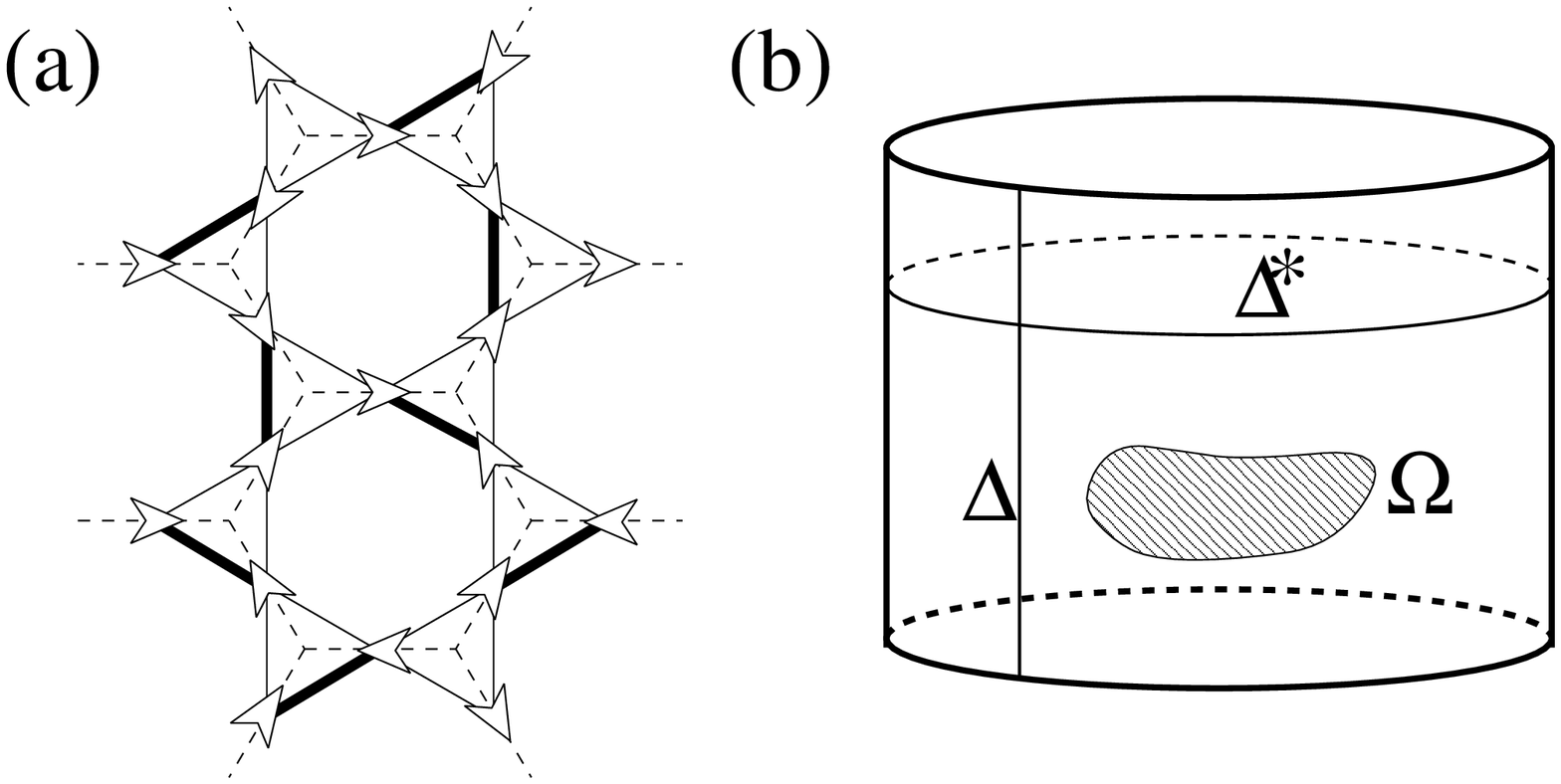}}
\caption{\label{figure_kagome}
(a) Dimer covering and arrow representation.  
(b)Local area $\Omega$ on the cylinder.
The cut $\Delta$ and the loop $\Delta^*$ can be drawn 
so as not to touch $\Omega$.}
\end{figure}

This model is simply expressed in terms of the arrow representation 
\cite{Elser} of dimer coverings defined in the following way.  
The sites of the kagome lattice $K$ can be identified with 
the centers of the bonds of the hexagonal lattice $H$.  
For a dimer covering of $K$, 
we assign orientations (arrows) to the bonds of $H$ 
so that the arrow on each site of $K$ 
points towards the interior of the triangle of $K$ 
where the dimer occupying the site is 
(see Fig. \ref{figure_kagome}(a)).  
As a consequence, the number of incoming arrows is even (0 or 2) 
at every triangle.  
Let $S$ be the set of arrow configurations satisfying 
this local parity constraint at every triangle.  
There is a one-to-one correspondence between $S$ and 
the set of all dimer coverings.  

We define $\tau^z(i)$ as the operator which flips the arrow 
on the site $i$ of K.  
Dimer movements can be represented as loop products of 
$\tau^z$ operators.  
The Hamiltonian we consider is the sum of the loop operators 
around the hexagons $h$ of $K$: 
$
 {\cal H}=-\sum_h \prod_{\alpha=1}^6 \tau^z(i_{h,\alpha}),
$
where $i_{h,\alpha}$ are the six sites of the hexagon $h$.  

If this model is defined on a surface with a {\em non-trivial topology} 
(cylinder, torus, etc.), 
arrow configurations in $S$ can be grouped into topological sectors 
which are not mixed by any succession of local dimer moves.  
From now on, 
we concentrate on the case of the cylinder for simplicity 
(but all the results can be easily generalized to other topologies).  
We draw a cut $\Delta$ (passing through the bonds of $H$) 
going from the top to the bottom of the cylinder.  
We classify arrow configurations into two topological sectors, 
$S^+$ and $S^-$, 
depending on whether the number of arrows crossing $\Delta$ 
to the right is even or odd.  
The spectrum can be determined separately in each sector.  
Using the standard Rokhsar-Kivelson argument \cite{Rokhsar}, 
one can show that 
the ground state in a given sector is exactly 
the equal-amplitude superposition of 
all dimer coverings (arrow configurations) belonging to that sector: 
\begin{equation}
 |\mu\rangle = 
 \frac{1}{\sqrt{|S^\mu|}} 
 \sum_{a\in S^\mu} |a\rangle,  \quad \mu=+,-.  
\end{equation}
These two states are exactly degenerate and 
form a 2-dimensional GS subspace.  

Now we consider the RDM of a state $\ket{\Psi}$ in the GS subspace 
and discuss how it depends on the choice of $\ket{\Psi}$. 
The area $\Omega$ is given as a set of bonds of $H$.  
The RDM $\rho_\Omega=\Tr_{\Omegabar}\ket{\Psi}\bra{\Psi}$ 
is defined by tracing out the degrees of freedom (arrows) on $\Omegabar$:
\begin{equation}
 \bra{a_1}\rho_\Omega\ket{a_2}
 = \sum_{\abar} \bracket{a_1,\abar}{\Psi} \bracket{\Psi}{a_2,\abar}, 
\end{equation}
where $a_1$ and $a_2$ are arrow configurations on $\Omega$ 
and the sum is over all the arrow configurations $\abar$ on $\Omegabar$.  
By expressing $|\Psi\rangle=\sum_\mu \alpha_\mu |\mu\rangle$ 
with $\sum_\mu |\alpha_\mu|^2=1$, 
$\rho_\Omega$ can be expanded as 
\begin{equation}\label{rho_expand}
 \rho_\Omega=\sum_{\mu,\nu} \alpha_\mu\alpha_\nu^* 
                            \Mcal_\Omega^{\mu\nu},\quad
 \Mcal_\Omega^{\mu\nu}=\Tr_{\Omegabar} |\mu\rangle\langle\nu|.  
\end{equation}

First we assume that $\Omega$ is a (finite) local area 
; see Fig. \ref{figure_kagome}(b).  
We prove the following relations: 
\begin{equation}\label{rdm_relations}
 \Mcal_\Omega^{+-}=0,\quad 
 \Mcal_\Omega^{\mu\nu}=\Mcal_\Omega^{\mubar\nubar},
\end{equation}
where the overbars represent the sign flip.  
To prove the first relation, we choose the cut $\Delta$ 
so as not to touch $\Omega$.  
Then the parity along $\Delta$ for an arrow configuration 
depends only on its part on $\Omegabar$.  
Let us consider the matrix element of $\Mcal_\Omega^{+-}$:  
$\bra{a_1}\Mcal_\Omega^{+-}\ket{a_2}
=\sum_{\abar} \bracket{a_1,\abar}{+} \bracket{-}{a_2,\abar}$.  
Since the two configurations, $(a_1,\abar)$ and $(a_2,\abar)$, 
have common parity, 
$\bracket{a_1,\abar}{+}$ and $\bracket{-}{a_2,\abar}$ 
cannot be non-zero at the same time 
and hence we obtain $\Mcal_\Omega^{+-}=0$.  

To prove the second relation in Eq. \eqref{rdm_relations}, 
we draw a loop $\Delta^*$ 
(passing along the bonds of $H$) encircling the cylinder 
so as not to touch the area $\Omega$.  
We introduce a loop operator along $\Delta^*$: 
$T_{\Delta^*}^z=\Pi_{i\in\Delta^*} \tau^z(i)$.  
This operator acts only on $\Omegabar$ and 
maps $\ket{\pm}$ to $\ket{\mp}$, showing 
$\Mcal_\Omega^{\mu\nu}
 =\Tr_{\Omegabar}\left(T_{\Delta^*}^z\ket{\mubar}
                      \bra{\nubar}T_{\Delta^*}^z\right)
 =\Tr_{\Omegabar}\left(\ket{\mubar}
                      \bra{\nubar}(T_{\Delta^*}^z)^2\right)
 =\Mcal_\Omega^{\mubar\nubar}$.  

From Eqs. \eqref{rho_expand} and \eqref{rdm_relations}, 
we see that $\rho_\Omega$ is independent of 
the choice of $|\Psi\rangle$. 
Thus no order parameter can be defined on an arbitrary local area.  
In the context of topological quantum bit based on QDMs
\cite{Misguich_qubit,Ioffe_Nature}, 
this shows the stability of quantum information 
against external noises (coupling locally to dimers).  

The situation is different 
if $\Omega$ has a non-trivial topology, namely, 
extends from the top to the bottom of the cylinder (case A)) 
or encircles the cylinder (case B)).  
In case A), we can choose the cut $\Delta$ inside $\Omega$.  
Then the parity along $\Delta$ 
can be expressed as the operator acting on $\Omega$ 
and distinguishes the different topological sectors.  
It can be considered as a non-local order parameter 
distinguishing $\ket{+}$ and $\ket{-}$.  
Similarly, in case B), 
two states $(\ket{+}\pm\ket{-})/\sqrt{2}$ 
are distinguished by the loop operator $\Tzs$ 
with $\Delta^*$ defined inside $\Omega$.  

We have shown that the GSs of this QDM 
cannot be distinguished by any local operator
but by non-local operators defined on areas 
with non-trivial topologies.  
We comment that 
a similar result has been shown without using RDMs 
by Ioffe and Feigel'man \cite{Ioffe_PRB} 
in their study on a related model.  
We stress, however,  that our formulation based on RDMs 
has an advantage in its generality.  
As demonstrated in the ladder model, 
it can be applied to models without exact solutions,  
by combining it with numerical calculation, for example.  

{\em Conclusions ---}
We have developed a method which can determine order parameters 
without using any empirical knowledge.  
The two applications confirmed 
its applicability to exotic orders 
and, especially,  its relevance for analyzing topological orders.  
We expect that our method will shed some light on the controversies 
in some frustrated quantum antiferromagnets 
(see Ref. \cite{Misguich_review} and references therein), 
e.g., the $J_1-J_2$ model on the square lattice 
and the multiple-spin exchange model on the triangular lattice.   

We are grateful to 
Vincent Pasquier, Masahiro Sato, and Didina Serban 
for many fruitful discussions.  
In performing numerical diagonalization, 
we benefited from TITPACK ver.2 
developed by H. Nishimori and LAPACK.  
This work was partially supported 
by the Grant-in-Aid for Scientific Research on Priority Areas, 
and by a 21st Century COE Program at Tokyo Tech
``Nanometer-Scale Quantum Physics'', both from MEXT of Japan.


\begin{thebibliography}{}
\bibitem{note_timereversal}
  This definition of the time-reversal operator 
  is different from the usual one accompanying the spin flip, 
  but is more convenient here 
  since the eigenstates obtained by Lanczos diagonalization are real.
\end{thebibliography}
\end{document}